\documentclass[%
reprint,
superscriptaddress,
nofootinbib,
 amsmath,amssymb,
 aps,
 prl
]{revtex4-1}

\usepackage{graphicx}	
\usepackage{dcolumn}	
\usepackage{bm} 		
\usepackage{paralist}
\usepackage[usenames,dvipsnames]{color}
\usepackage{float}
\usepackage{blindtext}
\usepackage[separate-uncertainty = true,multi-part-units=single]{siunitx}
\usepackage{suffix}

\newcommand*\figref[1]{Fig.\ \ref{#1}}
\newcommand*\ie{i.e.\ }
\newcommand*\eg{e.g.\ }

\DeclareMathOperator\erf{erf}

\begin{document}

\title{Point-projection microscopy of nano-localized photoemission currents at sub-40 femtosecond time scales}

\author{Faruk Kre\v{c}ini\'c}
\email{krecinic@fhi-berlin.mpg.de}
\author{Jannik Malter}
\author{Alexander Paarmann}
\author{Melanie M\"{u}ller}
\author{Ralph Ernstorfer}
\affiliation{Fritz-Haber-Institut der Max-Planck-Gesellschaft, Faradayweg 4-6, 14195 Berlin, Germany}

\date{\today}

\begin{abstract}
Femtosecond point-projection microscopy (fs-PPM) is an electron microscopy technique that possesses a combination of high spatio-temporal resolution and sensitivity to local electric fields. 
This allows it to visualize ultrafast charge carrier dynamics in complex nanomaterials. 
We benchmark the capability of the fs-PPM technique by imaging the ultrafast dynamics of charge carriers produced by multiphoton ionization of silver nanowires.
The space-charge driven motion of photoelectrons is followed on \mbox{sub-100}\,nm length scales, while the dynamics are captured on 30--100\,fs time scales.
The build-up of electron holes in the silver nanowires following photoelectron ejection, \ie positive charging, has also been observed.
The fastest observed photoelectron temporal response is 33\,fs (FWHM), which represents an upper estimate of the instrument response function and is consistent with an expected electron wavepacket duration of 13\,fs based on simulations.
\end{abstract}

\maketitle

The functionality of a nanomaterial is inherently linked to the composition and structure of its nanoscale building blocks, \eg quantum dots \cite{Nozik2010}, nanowires \cite{Hu1999} and/or 2D materials \cite{Geim2013}.
The interplay between structure and charge transport at the nanoscale in these complex systems often plays a crucial role in the development of new technological devices and applications.
Directly visualizing charge carrier motion in these systems requires a technique that ideally possesses femtosecond temporal resolution, nanometer spatial resolution, and a high sensitivity to excited charge carrier populations.
Over the past few decades powerful ultrafast microscopy techniques have been developed that have, to varying extents, realized this combination of extreme requirements.

Optical pump-probe microscopy and spectroscopy \cite{Grumstrup2015} have, for example, been extensively used to study charge carrier generation and propagation in many diverse and technologically relevant systems \cite{Gundlach2009,Grumstrup2014,Seo2016,Grancini2012,Guo2017}.
The temporal resolution of ultrafast optical microscopy is fundamentally only limited by the optical probe pulse duration, which can be well below 10\,fs.
However, the attainable spatial resolution is limited by the wavelength of light and is, consequently, on the order of a few hundred nanometer.
Using near-field confinement \cite{Kravtsov2016}, or decreasing the wavelength of the light into the XUV \cite{Zhang2015,Tadesse2016} or soft X-ray spectral range \cite{Fischer2006}, it is possible to improve the spatial resolution of optical pump-probe imaging techniques down to 10\,nm.

A fundamentally different approach is taken by ultrafast microscopy techniques that employ electrons as probes.
The inherently small de Broglie wavelength of electrons in principle enables achieving a spatial resolution well beyond that of optical microscopy techniques.
For example, time-resolved photoelectron emission microscopy has achieved spatial resolutions on the order of few tens of nm and a temporal resolution equivalent to that of optical microscopy techniques \cite{Kubo2005,MeyerzuHeringdorf2007,Man2016,Dabrowski2016}.
Techniques that seek to expand conventional scanning and transmission electron microscopy to the femtosecond regime and are capable of visualizing nanoscale carrier motion have also been demonstrated \cite{uelec:Zewail,Barwick2009,Najafi2015,Feist2016,Liao2017,DaSilva2017}.
A major challenge for the latter approach has been the dispersive and space-charge induced electron pulse spreading during propagation through the electron optical column of the microscope.
This currently limits the achievable temporal resolution to the few hundred femtosecond range, though techniques based on coherent electron-light interactions have been demonstrated that could enable electron microscopy with attosecond temporal resolution in the future \cite{Priebe2017,Morimoto2017}.

An attractive alternative for achieving nanometer spatial and femtosecond temporal resolution using electrons as probes is femtosecond point-projection microscopy (fs-PPM) \cite{Quinonez2013,Muller2014,Bainbridge2016}.
The combination of a nanotip electron source and lens-less imaging allows for a temporal resolution comparable to ultrafast optical microscopy and a spatial resolution that can, in principle, reach the single nm level \cite{Longchamp2015a,Paarmann2012,Vogelsang2015,Muller2016a}.
Due to the use of low-energy, \ie \mbox{sub-1\,kV}, electrons, fs-PPM is highly sensitive to local electric fields \cite{Beyer2010}, which makes it especially well suited to visualizing charge carrier separation and dynamics \cite{Muller2014}.
The first experimental studies with fs-PPM were applied to imaging surface photovoltage dynamics in semiconductor nanowires \cite{Muller2014} and photoelectron generation from nanotips \cite{Bainbridge2016}.
These initial experiments established the capability of the fs-PPM technique to reach \mbox{sub-100}\,nm spatial resolution, while imaging charge carrier dynamics on a 100\,fs timescale.
However, though simulations have indicated a temporal resolution comparable to the optical excitation pulse duration, so far the experiments employing fs-PPM have not demonstrated experimentally the capability of the technique to image \mbox{sub-100\,fs} dynamics and were either limited by the slow sample system dynamics \cite{Muller2014}, or by the large tip-sample distance \cite{Bainbridge2016} and the resulting long electron pulse durations.

In this paper, we report a benchmark experiment that visualizes dynamics that are considerably faster than what has been shown so far with this technique, illustrating the potential of fs-PPM in combining nanoscale spatial and femtosecond temporal resolution in imaging charge carrier dynamics.
The multiphoton ionization and subsequent space-charge driven dynamics of photoelectrons emitted from silver nanowires is captured on timescales as low as 33\,fs. 
It is also shown that the dynamics are highly dependent on the sample structure and that they vary on a \mbox{sub-100}\,nm scale, demonstrating the advantage of combining nanometer spatial and femtosecond temporal resolution.

\begin{figure}
\includegraphics[width=\columnwidth]{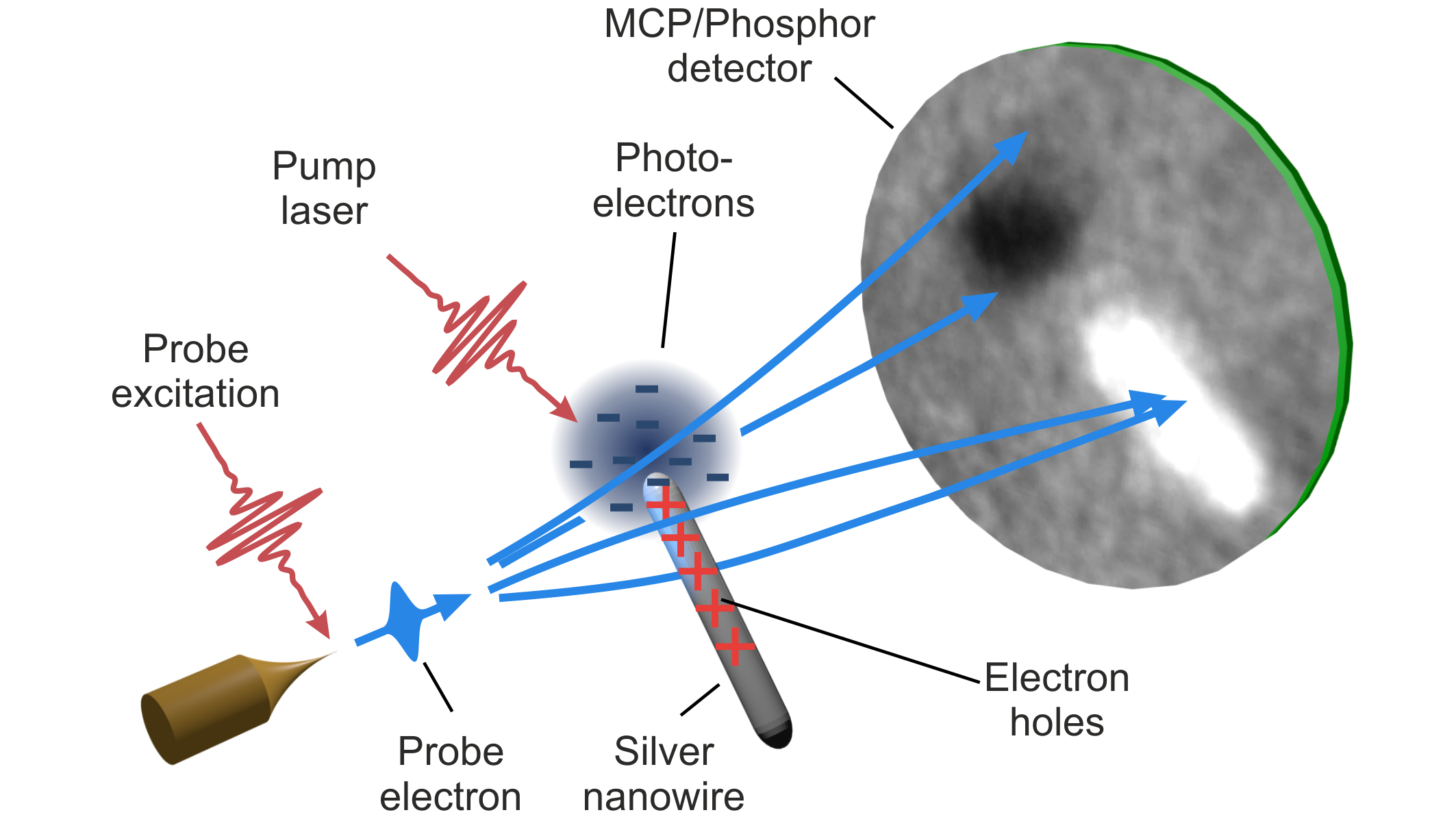}
\caption{Schematic illustration of the experimental setup. A nanotip that acts as a point-source of ultrafast electron pulses is brought close to the sample such that the diverging electron beam projects a magnified image of the sample onto the detector. 
Local fields, due to \eg photoelectrons or electron holes, can deflect the probe electrons and result in bright and dark regions in the projected image.}
\label{fig:expsetup}
\end{figure}

A schematic illustration of the experimental setup is shown in \figref{fig:expsetup}. 
In PPM a nanotip that acts as a point-like source of electrons is brought close to the sample such that the diverging electron beam projects a magnified image of the sample onto the detector \cite{Morton1939,Spence1993}.
In fs-PPM the imaging electrons are generated by focusing an ultrashort laser pulse on the apex of a metallic nanotip, which leads to non-linear photoemission of an ultrashort electron pulse.
The combination of a nanotip source and lens-less imaging allows for an ultra-compact microscope column, where the distance from the electron source to the sample is on the micrometer scale.
This minimizes the effect of dispersive broadening and allows for an electron pulse duration comparable to that of the optical excitation pulse \cite{Paarmann2012}. 
It is important to note that the projection image produced by the electron beam is not merely a geometric projection of the sample, but contains information on electric fields at the sample through lensing effects \cite{Beyer2010,Lai1999}.
Local fields cause the imaging (probe) electrons to be deflected while passing by the sample and result in the appearance of bright or dark features in the projected image, depending on whether the electrons are attracted or repelled by those fields (see \figref{fig:expsetup}).

The experiment was performed using a pump-probe scheme, in which an ultrashort laser pulse excites the sample, photoionizing the silver nanowires, and a time-delayed electron pulse records an ultrafast image of the nanoscale system.
Laser pulses of 8\,fs duration at 800\,nm and a repetition rate of 1 MHz were used for the experiment, with pump and probe excitation pulse energies of \SI{10}{n J} and \SI{0.6}{n J}, respectively.
Both laser pulses were focused onto the sample and tip apex by silver parabolic mirrors, reaching estimated fluences of \SI{14}{mJ/cm^2} and \SI{2}{mJ/cm^2}, respectively.
In order to facilitate optical access of the probe excitation laser pulse to the nanotip the sample was inclined with respect to the nanotip by $\sim$\ang{8.5}.
The imaging electrons were accelerated by a tip voltage of -207 V, while the sample was kept at ground potential.
The tip-sample distance was \SI{20}{\micro m}, resulting in a geometric magnification factor of $\sim$5000, and was chosen such that there was no observable photoelectron generation from the nanotip by the pump pulse, while at the same time minimizing the propagation distance for the electron pulse. 
Silver nanowires (Sigma-Aldrich) with a nominal diameter of \SI{60}{nm} and a length of \SI{10}{\micro m} were deposited from a colloidal solution onto a lacey carbon transmission electron microscopy grid by dropcasting.


\begin{figure*}
\includegraphics[width=\textwidth]{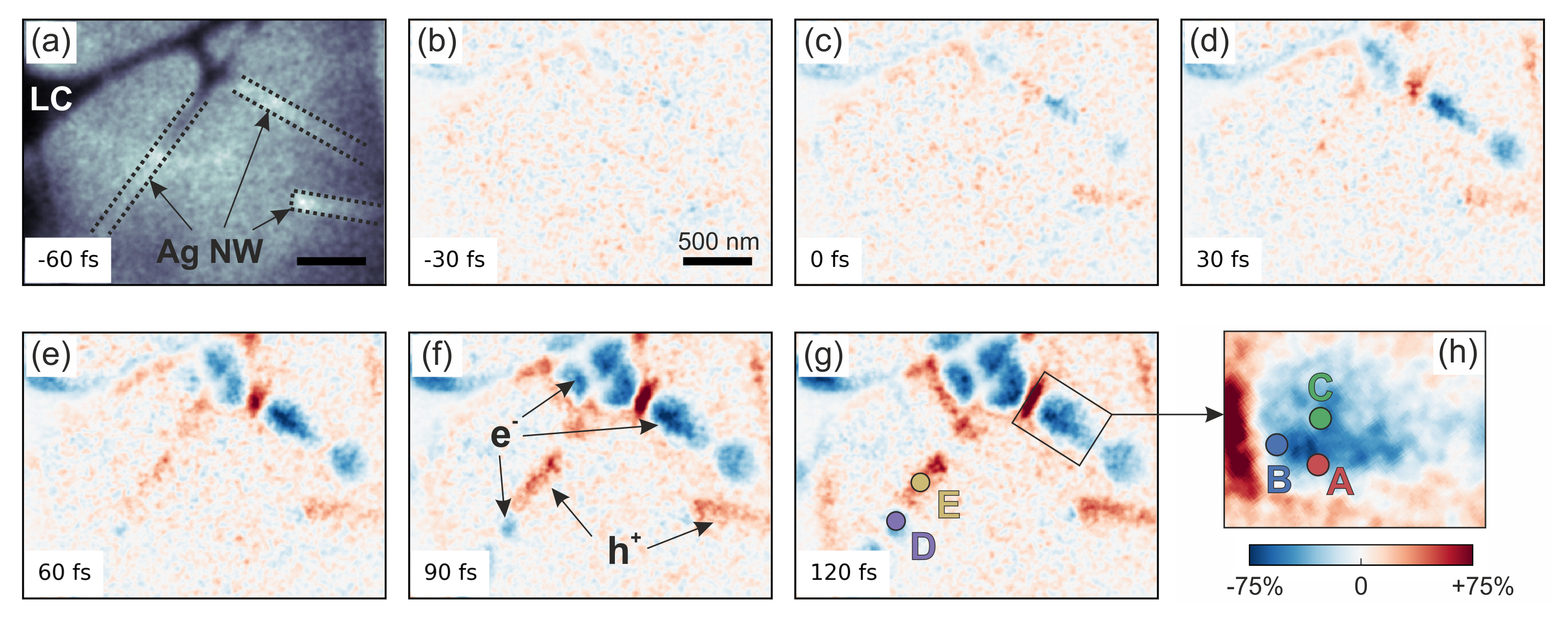}
\caption{Time sequence of multiphoton ionization of silver nanowires in 30\,fs increments of the pump-probe delay time: (a) reference image of the sample before photoexcitation where silver nanowires (Ag NW) and the lacey carbon (LC) sample support is indicated; (b)--(g) difference images with respect to (a), in (f) photoelectron emission and nanowire image charge effects are indicated by (e$^-$) and (h$^+$), respectively; (h) enlarged image of the specific photoelectron emission region that is analyzed in \figref{fig:fitmapI}. Scale bars: \SI{500}{nm}.}
\label{fig:tseq}
\end{figure*}

\begin{figure}
\centering
\includegraphics[width=0.8\columnwidth]{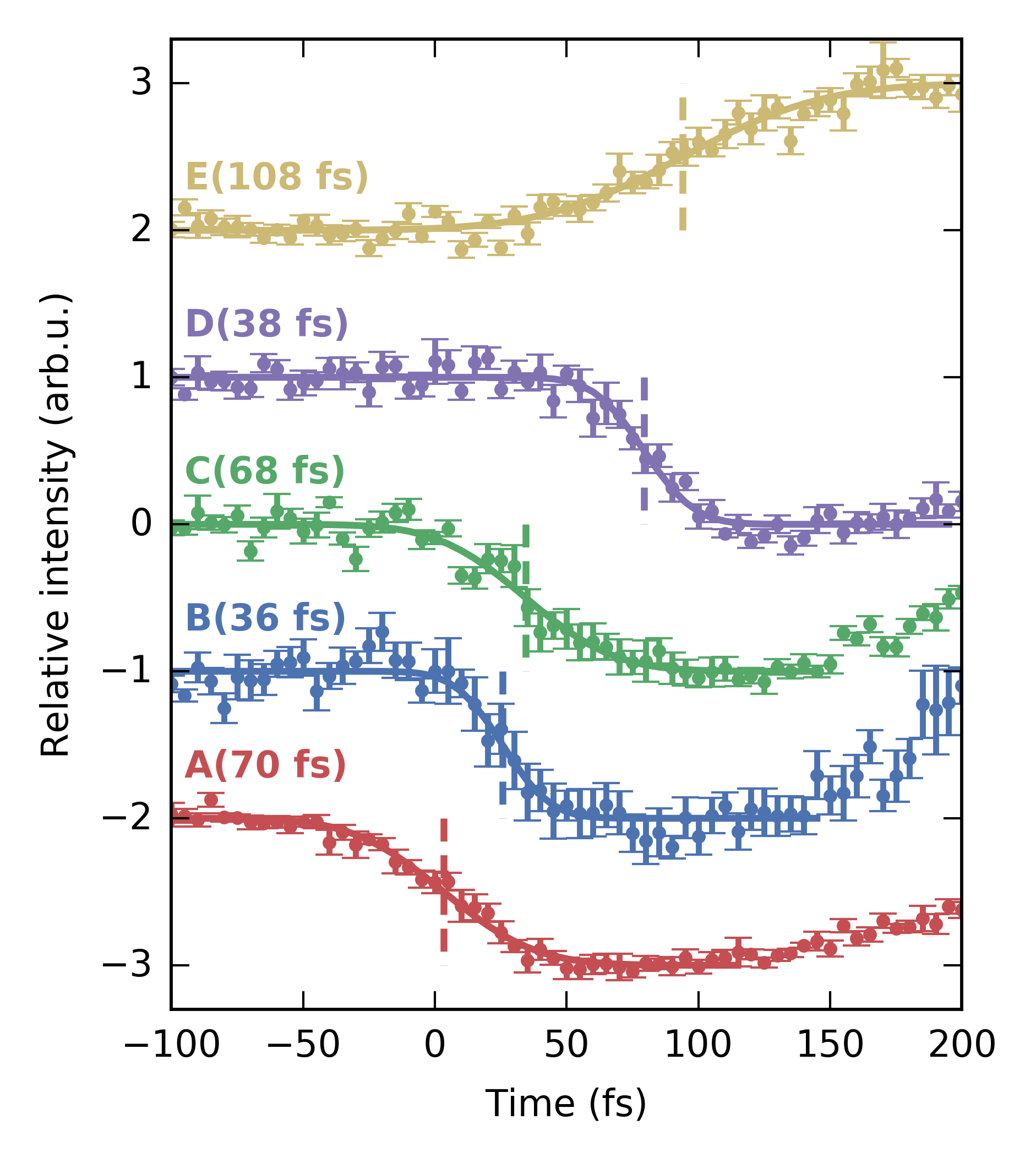}
\caption{Time responses at points A--E indicated in \figref{fig:tseq}(g) and (h), and the corresponding error-function fits (solid lines). The full width at half maximum time of the Gaussian distribution corresponding to the fitted curves are indicated in brackets and the time zero positions by dashed lines.}
\label{fig:XCsites}
\end{figure}

Figure \ref{fig:tseq} shows an experimental pump-probe time-sequence, where frames (b)--(g) show difference images with respect to the reference image (a) of the sample before photoexcitation.
The silver nanowires are visible in the reference image as bright stripes, due to the presence of static electric fields that lead to the formation of an electron biprism lens \cite{Beyer2010,Lai1999}.
After photoexcitation, locations with strong photoelectron emission are clearly visible in the difference images as regions with transiently reduced transmission, \eg blue color regions annotated with e$^-$ in \figref{fig:tseq}(f).
In this case the presence of the photoelectrons leads to a locally repelling electric field on the imaging electrons that acts as a negative lens and causes a reduction of up to 75\% of the intensity  in the projected image.
Besides the strong transient suppression of the signal in regions where photoelectrons are emitted, the positive charging of the nanowires themselves can be observed as well, \eg red color regions indicated by h$^+$ in \figref{fig:tseq}(f).
The positive charging of the nanowires upon emission of photoelectrons leads to an enhancement of the electron biprism effect and therefore to an increase in their apparent brightness in the projection images.
Finally, we point out that in the difference image time-sequence the dynamics in the top right corner of the images appears to start sooner than in the lower left corner. 
In \figref{fig:tseq}(c) clear photoelectron emission signatures can be seen in the top right corner, whereas the first onset of photoelectron emission in the lower left corner only becomes visible in frame (e), after a $\sim$60\,fs delay. 
This effect can be attributed to propagation path length differences that result from the \ang{8.5} inclination of the sample with respect to the probing electron beam, and the geometric curvature of the electron probe pulses produced by the the point-like electron source \cite{Paarmann2012}.
Due to the low velocity of the probe electrons there can be a considerable time delay in the arrival of the probe pulse across the sample, which causes an effective shift of the time zero across the field-of-view.

We now investigate the dynamic behavior at a few exemplary positions in more detail, labeled as points A--E in \figref{fig:tseq}(g) and (h).
The time response at these points is shown in \figref{fig:XCsites}, together with fitted  error-function curves (solid lines) of the form $I(t)=A + B(1+\erf((t-t_0)/\sqrt{2}\sigma)$.
For each curve the full width at half maximum (FWHM) time $\tau$ of the Gaussian distribution corresponding to the fitted standard deviation $\sigma$ is indicated in brackets, while the fitted time zero $t_0$ is marked by a vertical dashed line.
First, we observe that the time constants show a rather large spread in values: at points B and D we find $\tau < 40$\,fs, whereas at points A, C and E it varies between $64-99$\,fs.
In particular, we observe that the positively charging nanowires (point E) have a dynamic response that is much slower than that of the photoelectrons.
This is because the former involves hole propagation within the nanowire, whereas the latter is due to the propagation of photoelectrons driven by strong space-charge forces resulting from the high photoemission densities under the current experimental conditions.
Second, we see a clear shift in the fitted time zero at each position.
These time zero shifts can be attributed to two distinct effects: (1) probe electron path length differences, which were pointed out previously, and (2) charge carrier propagation effects.
The  $\sim$80\,fs time zero shift between points A and D, which are \SI{1.6}{\micro m} apart in the sample plane, can largely be explained by the first effect. 
However, points A, B and C are separated by $\sim$100\,nm from each other, such that probe path length difference effects can only account for a time zero difference of less than 4\,fs.
The time zero shifts between curves A, B and C, which are on the order of few tens of fs, can therefore be attributed to photoelectron propagation effects. 

\begin{figure}
\centering
\includegraphics[width=0.8\columnwidth]{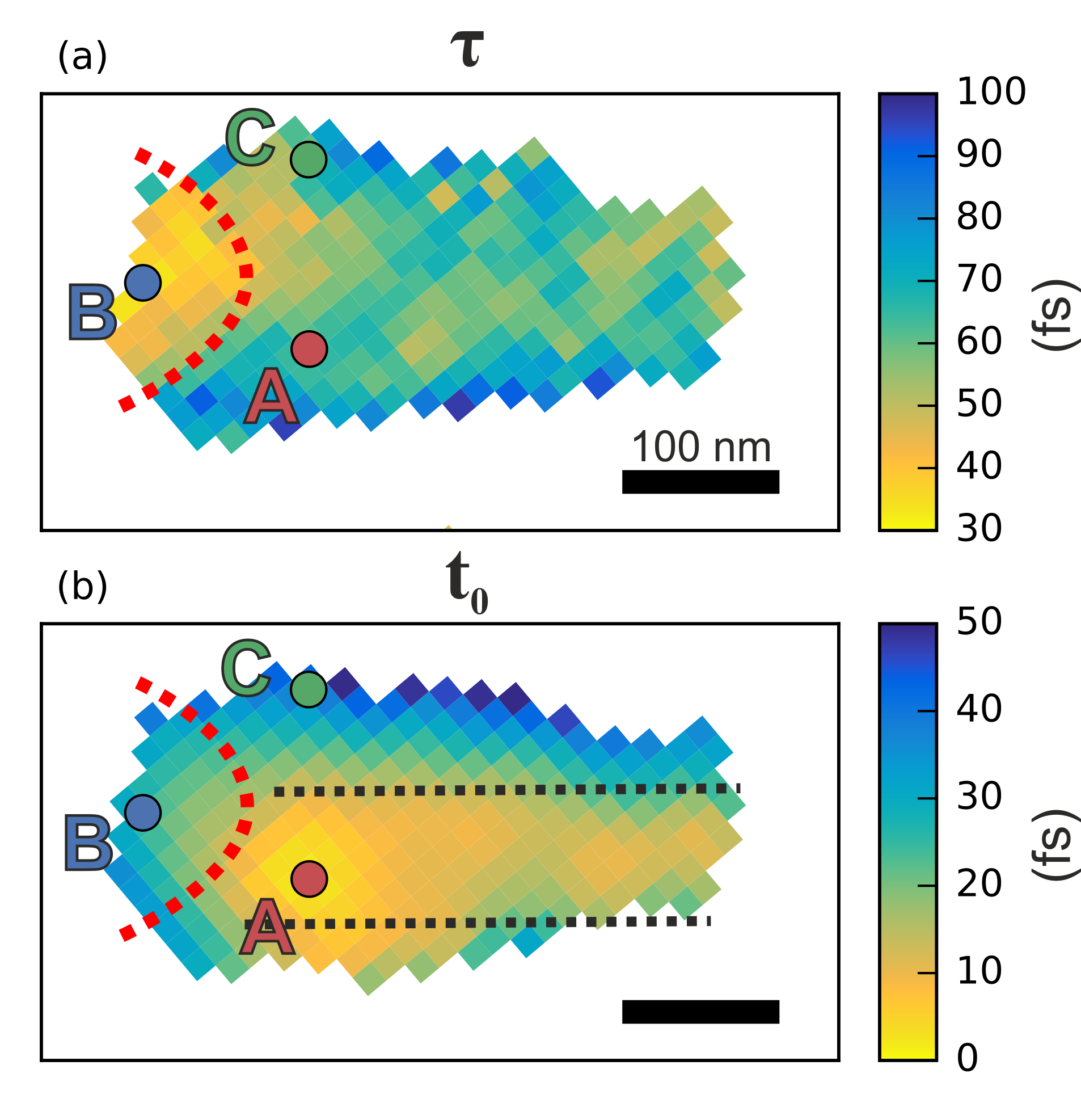}
\caption{Mapping of time constants (a) and time zeros (b) extracted from error-function fits in the region indicated in \figref{fig:tseq}(h). 
Scale bar: \SI{100}{nm}.}
\label{fig:fitmapI}
\end{figure}

In order to study photoelectron emission and propagation dynamics in the immediate vicinity of the nanowires we show in \figref{fig:fitmapI} a map of the time constants ($\tau$) and time zeros ($t_0$) for the entire region indicated in \figref{fig:tseq}(h).
The fitting constants are obtained from error-function fits, such as those shown in \figref{fig:XCsites}, at a regular grid within in the region of interest.
In the time constant map in \figref{fig:fitmapI}(a) we indeed see that the speed of the photoelectron dynamics is spatially inhomogeneous; the time constants of the dynamics in the small 100\,nm region to the left of the dashed red line are 30--40\,fs, while in the region to the right they are in the range of 40--100\,fs.
Furthermore, the expansion and propagation of the photoelectron charge distribution can be observed directly in the time zero map in \figref{fig:fitmapI}(b).
The position of the nanowire itself is visible as a rather uniform region with very low time zero values of 0--15\,fs located between the black dashed lines. 
Outside of this region there is a clear gradient in the extracted time zero values, with values of 15--50\,fs, that is attributed to the space-charge induced propagation of the photoelectrons away from the emission point at the nanowire. 
This highly inhomogeneous nature of photoelectron dynamics taking place on \mbox{sub-100}\,nm length and \mbox{sub-40}\,fs time scales clearly illustrates the usefulness of the combination of high temporal and spatial resolution realized with the fs-PPM technique.

Finally, we address the relation between the temporal resolution and the observed photoelectron dynamics in the current experiment.
Space-charge effects are frequently used as a cross-correlation tool for evaluating the temporal resolution of femtosecond electron diffraction and microscopy techniques \cite{Dolocan2006,uelec:Dwyer,Hebeisen2008,Centurion2008,Bainbridge2016}. 
Disregarding intrinsic photoelectron emission dynamics, the lowest time constants observed in the current experiment provide us with an upper limit on the instrument response function of 33\,fs~(FWHM).
In the single electron per pulse regime the fs-PPM setup was operated for this experiment, the probe electron pulse duration at the sample is determined by dispersive pulse broadening only.
An estimate of the dispersive pulse broadening for current experimental conditions was obtained using a particle tracing model that simulates the spread in arrival times of on-axis probe electrons with different initial energies \cite{Muller2016a,Paarmann2012}. 
Assuming an initial energy spread of 0.75 eV \cite{Quinonez2013,Yanagisawa2011} we obtain a pulse duration at the sample of 13\,fs~(FWHM).
Another important factor that affects the effective time resolution of the fs-PPM setup is the traversal time of the low energy probe electrons through the interaction volume in the sample plane.
Two factors that influence this are the sample system extent in the electron beam propagation direction ($\sim$60\,nm) and mechanical vibrations between the tip and sample ($\sim$100\,nm), which contribute an estimated additional 12\,fs to the effective temporal resolution.
Increasing the mechanical stability of the setup is therefore expected to improve both the temporal and spatial resolution.

In conclusion, the results presented here demonstrate the potential of the fs-PPM technique to image ultrafast charge carrier dynamics in a nanoscale system down to 30\,fs time and \mbox{sub-100}\,nm length scales. 
We expect that improving the mechanical stability of the experimental setup and using surface plasmon polariton based electron sources \cite{Vogelsang2015,Muller2016a} will lead to the realization of femtosecond low-energy electron holography (fs-LEEH) with few nm and sub-10\,fs spatio-temporal resolution.
Static low-energy electron holography has recently been shown to be extremely sensitive to local electric fields and is capable of imaging a single elementary charge adsorbed on graphene \cite{Latychevskaia2016}.
With fs-LEEH more detailed studies of the effects that influence photoelectron dynamics following photoemission, such as \eg electron re-scattering \cite{Yanagisawa2016}, will become possible.
Moreover, combining the high spatio-temporal resolution of fs-LEEH with holographic phase information may also allow for studying coherent plasmonic effects and electron-light interactions in near fields \cite{Barwick2009}.
More generally, we expect fs-LEEH to be especially suitable for the study of photoinduced charge carrier separation and motion in complex nano-structured materials and systems, such as quantum dots \cite{Zhang2015} nanowires \cite{Kinzel2016}, and 2D materials \cite{Prechtel2011,Hong2014}.
\smallskip
\begin{acknowledgments}
This project was funded by the Max Planck Society and by the European Research Council under the EU Horizon 2020 research and innovation program (Grant Agreement No.~ERC-2015-CoG-682843).
\end{acknowledgments}

\bibliography{Bibliography}{}

\begin{thebibliography}{46}%
\makeatletter
\providecommand \@ifxundefined [1]{%
 \@ifx{#1\undefined}
}%
\providecommand \@ifnum [1]{%
 \ifnum #1\expandafter \@firstoftwo
 \else \expandafter \@secondoftwo
 \fi
}%
\providecommand \@ifx [1]{%
 \ifx #1\expandafter \@firstoftwo
 \else \expandafter \@secondoftwo
 \fi
}%
\providecommand \natexlab [1]{#1}%
\providecommand \enquote  [1]{``#1''}%
\providecommand \bibnamefont  [1]{#1}%
\providecommand \bibfnamefont [1]{#1}%
\providecommand \citenamefont [1]{#1}%
\providecommand \href@noop [0]{\@secondoftwo}%
\providecommand \href [0]{\begingroup \@sanitize@url \@href}%
\providecommand \@href[1]{\@@startlink{#1}\@@href}%
\providecommand \@@href[1]{\endgroup#1\@@endlink}%
\providecommand \@sanitize@url [0]{\catcode `\\12\catcode `\$12\catcode
  `\&12\catcode `\#12\catcode `\^12\catcode `\_12\catcode `\%12\relax}%
\providecommand \@@startlink[1]{}%
\providecommand \@@endlink[0]{}%
\providecommand \url  [0]{\begingroup\@sanitize@url \@url }%
\providecommand \@url [1]{\endgroup\@href {#1}{\urlprefix }}%
\providecommand \urlprefix  [0]{URL }%
\providecommand \Eprint [0]{\href }%
\providecommand \doibase [0]{http://dx.doi.org/}%
\providecommand \selectlanguage [0]{\@gobble}%
\providecommand \bibinfo  [0]{\@secondoftwo}%
\providecommand \bibfield  [0]{\@secondoftwo}%
\providecommand \translation [1]{[#1]}%
\providecommand \BibitemOpen [0]{}%
\providecommand \bibitemStop [0]{}%
\providecommand \bibitemNoStop [0]{.\EOS\space}%
\providecommand \EOS [0]{\spacefactor3000\relax}%
\providecommand \BibitemShut  [1]{\csname bibitem#1\endcsname}%
\let\auto@bib@innerbib\@empty
\bibitem [{\citenamefont {Nozik}\ \emph {et~al.}(2010)\citenamefont {Nozik},
  \citenamefont {Beard}, \citenamefont {Luther}, \citenamefont {Law},
  \citenamefont {Ellingson},\ and\ \citenamefont {Johnson}}]{Nozik2010}%
  \BibitemOpen
  \bibfield  {author} {\bibinfo {author} {\bibfnamefont {A.~J.}\ \bibnamefont
  {Nozik}}, \bibinfo {author} {\bibfnamefont {M.~C.}\ \bibnamefont {Beard}},
  \bibinfo {author} {\bibfnamefont {J.~M.}\ \bibnamefont {Luther}}, \bibinfo
  {author} {\bibfnamefont {M.}~\bibnamefont {Law}}, \bibinfo {author}
  {\bibfnamefont {R.~J.}\ \bibnamefont {Ellingson}}, \ and\ \bibinfo {author}
  {\bibfnamefont {J.~C.}\ \bibnamefont {Johnson}},\ }\href
  {http://pubs.acs.org/doi/abs/10.1021/cr900289f} {\bibfield  {journal}
  {\bibinfo  {journal} {Chem. Rev.}\ }\textbf {\bibinfo {volume} {110}},\
  \bibinfo {pages} {6873} (\bibinfo {year} {2010})}\BibitemShut {NoStop}%
\bibitem [{\citenamefont {Hu}\ \emph {et~al.}(1999)\citenamefont {Hu},
  \citenamefont {Odom},\ and\ \citenamefont {Lieber}}]{Hu1999}%
  \BibitemOpen
  \bibfield  {author} {\bibinfo {author} {\bibfnamefont {J.}~\bibnamefont
  {Hu}}, \bibinfo {author} {\bibfnamefont {T.~W.}\ \bibnamefont {Odom}}, \ and\
  \bibinfo {author} {\bibfnamefont {C.~M.}\ \bibnamefont {Lieber}},\ }\href
  {http://pubs.acs.org/doi/abs/10.1021/ar9700365} {\bibfield  {journal}
  {\bibinfo  {journal} {Acc. Chem. Res.}\ }\textbf {\bibinfo {volume} {32}},\
  \bibinfo {pages} {435} (\bibinfo {year} {1999})}\BibitemShut {NoStop}%
\bibitem [{\citenamefont {Geim}\ and\ \citenamefont
  {Grigorieva}(2013)}]{Geim2013}%
  \BibitemOpen
  \bibfield  {author} {\bibinfo {author} {\bibfnamefont {A.~K.}\ \bibnamefont
  {Geim}}\ and\ \bibinfo {author} {\bibfnamefont {I.~V.}\ \bibnamefont
  {Grigorieva}},\ }\href {http://www.nature.com/doifinder/10.1038/nature12385}
  {\bibfield  {journal} {\bibinfo  {journal} {Nature}\ }\textbf {\bibinfo
  {volume} {499}},\ \bibinfo {pages} {419} (\bibinfo {year}
  {2013})}\BibitemShut {NoStop}%
\bibitem [{\citenamefont {Grumstrup}\ \emph {et~al.}(2015)\citenamefont
  {Grumstrup}, \citenamefont {Gabriel}, \citenamefont {Cating}, \citenamefont
  {{Van Goethem}},\ and\ \citenamefont {Papanikolas}}]{Grumstrup2015}%
  \BibitemOpen
  \bibfield  {author} {\bibinfo {author} {\bibfnamefont {E.~M.}\ \bibnamefont
  {Grumstrup}}, \bibinfo {author} {\bibfnamefont {M.~M.}\ \bibnamefont
  {Gabriel}}, \bibinfo {author} {\bibfnamefont {E.~E.}\ \bibnamefont {Cating}},
  \bibinfo {author} {\bibfnamefont {E.~M.}\ \bibnamefont {{Van Goethem}}}, \
  and\ \bibinfo {author} {\bibfnamefont {J.~M.}\ \bibnamefont {Papanikolas}},\
  }\href {http://linkinghub.elsevier.com/retrieve/pii/S0301010415001901}
  {\bibfield  {journal} {\bibinfo  {journal} {Chem. Phys.}\ }\textbf {\bibinfo
  {volume} {458}},\ \bibinfo {pages} {30} (\bibinfo {year} {2015})}\BibitemShut
  {NoStop}%
\bibitem [{\citenamefont {Gundlach}\ and\ \citenamefont
  {Piotrowiak}(2009)}]{Gundlach2009}%
  \BibitemOpen
  \bibfield  {author} {\bibinfo {author} {\bibfnamefont {L.}~\bibnamefont
  {Gundlach}}\ and\ \bibinfo {author} {\bibfnamefont {P.}~\bibnamefont
  {Piotrowiak}},\ }\href {http://pubs.acs.org/doi/abs/10.1021/jp9013509}
  {\bibfield  {journal} {\bibinfo  {journal} {J. Phys. Chem. C}\ }\textbf
  {\bibinfo {volume} {113}},\ \bibinfo {pages} {12162} (\bibinfo {year}
  {2009})}\BibitemShut {NoStop}%
\bibitem [{\citenamefont {Grumstrup}\ \emph {et~al.}(2014)\citenamefont
  {Grumstrup}, \citenamefont {Gabriel}, \citenamefont {Pinion}, \citenamefont
  {Parker}, \citenamefont {Cahoon},\ and\ \citenamefont
  {Papanikolas}}]{Grumstrup2014}%
  \BibitemOpen
  \bibfield  {author} {\bibinfo {author} {\bibfnamefont {E.~M.}\ \bibnamefont
  {Grumstrup}}, \bibinfo {author} {\bibfnamefont {M.~M.}\ \bibnamefont
  {Gabriel}}, \bibinfo {author} {\bibfnamefont {C.~W.}\ \bibnamefont {Pinion}},
  \bibinfo {author} {\bibfnamefont {J.~K.}\ \bibnamefont {Parker}}, \bibinfo
  {author} {\bibfnamefont {J.~F.}\ \bibnamefont {Cahoon}}, \ and\ \bibinfo
  {author} {\bibfnamefont {J.~M.}\ \bibnamefont {Papanikolas}},\ }\href
  {http://pubs.acs.org/doi/abs/10.1021/nl5026166} {\bibfield  {journal}
  {\bibinfo  {journal} {Nano Lett.}\ }\textbf {\bibinfo {volume} {14}},\
  \bibinfo {pages} {6287} (\bibinfo {year} {2014})}\BibitemShut {NoStop}%
\bibitem [{\citenamefont {Seo}\ \emph {et~al.}(2016)\citenamefont {Seo},
  \citenamefont {Yamaguchi}, \citenamefont {Mohite}, \citenamefont
  {Boubanga-Tombet}, \citenamefont {Blancon}, \citenamefont {Najmaei},
  \citenamefont {Ajayan}, \citenamefont {Lou}, \citenamefont {Taylor},\ and\
  \citenamefont {Prasankumar}}]{Seo2016}%
  \BibitemOpen
  \bibfield  {author} {\bibinfo {author} {\bibfnamefont {M.}~\bibnamefont
  {Seo}}, \bibinfo {author} {\bibfnamefont {H.}~\bibnamefont {Yamaguchi}},
  \bibinfo {author} {\bibfnamefont {A.~D.}\ \bibnamefont {Mohite}}, \bibinfo
  {author} {\bibfnamefont {S.}~\bibnamefont {Boubanga-Tombet}}, \bibinfo
  {author} {\bibfnamefont {J.-C.}\ \bibnamefont {Blancon}}, \bibinfo {author}
  {\bibfnamefont {S.}~\bibnamefont {Najmaei}}, \bibinfo {author} {\bibfnamefont
  {P.~M.}\ \bibnamefont {Ajayan}}, \bibinfo {author} {\bibfnamefont
  {J.}~\bibnamefont {Lou}}, \bibinfo {author} {\bibfnamefont {A.~J.}\
  \bibnamefont {Taylor}}, \ and\ \bibinfo {author} {\bibfnamefont {R.~P.}\
  \bibnamefont {Prasankumar}},\ }\href
  {http://www.nature.com/articles/srep21601} {\bibfield  {journal} {\bibinfo
  {journal} {Sci. Rep.}\ }\textbf {\bibinfo {volume} {6}},\ \bibinfo {pages}
  {21601} (\bibinfo {year} {2016})}\BibitemShut {NoStop}%
\bibitem [{\citenamefont {Grancini}\ \emph {et~al.}(2012)\citenamefont
  {Grancini}, \citenamefont {Biasiucci}, \citenamefont {Mastria}, \citenamefont
  {Scotognella}, \citenamefont {Tassone}, \citenamefont {Polli}, \citenamefont
  {Gigli},\ and\ \citenamefont {Lanzani}}]{Grancini2012}%
  \BibitemOpen
  \bibfield  {author} {\bibinfo {author} {\bibfnamefont {G.}~\bibnamefont
  {Grancini}}, \bibinfo {author} {\bibfnamefont {M.}~\bibnamefont {Biasiucci}},
  \bibinfo {author} {\bibfnamefont {R.}~\bibnamefont {Mastria}}, \bibinfo
  {author} {\bibfnamefont {F.}~\bibnamefont {Scotognella}}, \bibinfo {author}
  {\bibfnamefont {F.}~\bibnamefont {Tassone}}, \bibinfo {author} {\bibfnamefont
  {D.}~\bibnamefont {Polli}}, \bibinfo {author} {\bibfnamefont
  {G.}~\bibnamefont {Gigli}}, \ and\ \bibinfo {author} {\bibfnamefont
  {G.}~\bibnamefont {Lanzani}},\ }\href
  {http://pubs.acs.org/doi/10.1021/jz3000382} {\bibfield  {journal} {\bibinfo
  {journal} {J. Phys. Chem. Lett.}\ }\textbf {\bibinfo {volume} {3}},\ \bibinfo
  {pages} {517} (\bibinfo {year} {2012})}\BibitemShut {NoStop}%
\bibitem [{\citenamefont {Guo}\ \emph {et~al.}(2017)\citenamefont {Guo},
  \citenamefont {Wan}, \citenamefont {Yang}, \citenamefont {Snaider},
  \citenamefont {Zhu},\ and\ \citenamefont {Huang}}]{Guo2017}%
  \BibitemOpen
  \bibfield  {author} {\bibinfo {author} {\bibfnamefont {Z.}~\bibnamefont
  {Guo}}, \bibinfo {author} {\bibfnamefont {Y.}~\bibnamefont {Wan}}, \bibinfo
  {author} {\bibfnamefont {M.}~\bibnamefont {Yang}}, \bibinfo {author}
  {\bibfnamefont {J.}~\bibnamefont {Snaider}}, \bibinfo {author} {\bibfnamefont
  {K.}~\bibnamefont {Zhu}}, \ and\ \bibinfo {author} {\bibfnamefont
  {L.}~\bibnamefont {Huang}},\ }\href
  {http://www.sciencemag.org/lookup/doi/10.1126/science.aam7744} {\bibfield
  {journal} {\bibinfo  {journal} {Science}\ }\textbf {\bibinfo {volume}
  {356}},\ \bibinfo {pages} {59} (\bibinfo {year} {2017})}\BibitemShut
  {NoStop}%
\bibitem [{\citenamefont {Kravtsov}\ \emph {et~al.}(2016)\citenamefont
  {Kravtsov}, \citenamefont {Ulbricht}, \citenamefont {Atkin},\ and\
  \citenamefont {Raschke}}]{Kravtsov2016}%
  \BibitemOpen
  \bibfield  {author} {\bibinfo {author} {\bibfnamefont {V.}~\bibnamefont
  {Kravtsov}}, \bibinfo {author} {\bibfnamefont {R.}~\bibnamefont {Ulbricht}},
  \bibinfo {author} {\bibfnamefont {J.~M.}\ \bibnamefont {Atkin}}, \ and\
  \bibinfo {author} {\bibfnamefont {M.~B.}\ \bibnamefont {Raschke}},\ }\href
  {http://www.nature.com/doifinder/10.1038/nnano.2015.336} {\bibfield
  {journal} {\bibinfo  {journal} {Nat. Nanotechnol.}\ }\textbf {\bibinfo
  {volume} {11}},\ \bibinfo {pages} {459} (\bibinfo {year} {2016})}\BibitemShut
  {NoStop}%
\bibitem [{\citenamefont {Zhang}\ \emph {et~al.}(2015)\citenamefont {Zhang},
  \citenamefont {Chernomordik}, \citenamefont {Crisp}, \citenamefont {Kroupa},
  \citenamefont {Luther}, \citenamefont {Miller}, \citenamefont {Gao},\ and\
  \citenamefont {Beard}}]{Zhang2015}%
  \BibitemOpen
  \bibfield  {author} {\bibinfo {author} {\bibfnamefont {J.}~\bibnamefont
  {Zhang}}, \bibinfo {author} {\bibfnamefont {B.~D.}\ \bibnamefont
  {Chernomordik}}, \bibinfo {author} {\bibfnamefont {R.~W.}\ \bibnamefont
  {Crisp}}, \bibinfo {author} {\bibfnamefont {D.~M.}\ \bibnamefont {Kroupa}},
  \bibinfo {author} {\bibfnamefont {J.~M.}\ \bibnamefont {Luther}}, \bibinfo
  {author} {\bibfnamefont {E.~M.}\ \bibnamefont {Miller}}, \bibinfo {author}
  {\bibfnamefont {J.}~\bibnamefont {Gao}}, \ and\ \bibinfo {author}
  {\bibfnamefont {M.~C.}\ \bibnamefont {Beard}},\ }\href
  {http://pubs.acs.org/doi/abs/10.1021/acsnano.5b01859} {\bibfield  {journal}
  {\bibinfo  {journal} {ACS Nano}\ }\textbf {\bibinfo {volume} {9}},\ \bibinfo
  {pages} {7151} (\bibinfo {year} {2015})}\BibitemShut {NoStop}%
\bibitem [{\citenamefont {Tadesse}\ \emph {et~al.}(2016)\citenamefont
  {Tadesse}, \citenamefont {Klas}, \citenamefont {Demmler}, \citenamefont
  {H{\"{a}}drich}, \citenamefont {Wahyutama}, \citenamefont {Steinert},
  \citenamefont {Spielmann}, \citenamefont {Z{\"{u}}rch}, \citenamefont
  {Pertsch}, \citenamefont {T{\"{u}}nnermann}, \citenamefont {Limpert},\ and\
  \citenamefont {Rothhardt}}]{Tadesse2016}%
  \BibitemOpen
  \bibfield  {author} {\bibinfo {author} {\bibfnamefont {G.~K.}\ \bibnamefont
  {Tadesse}}, \bibinfo {author} {\bibfnamefont {R.}~\bibnamefont {Klas}},
  \bibinfo {author} {\bibfnamefont {S.}~\bibnamefont {Demmler}}, \bibinfo
  {author} {\bibfnamefont {S.}~\bibnamefont {H{\"{a}}drich}}, \bibinfo {author}
  {\bibfnamefont {I.}~\bibnamefont {Wahyutama}}, \bibinfo {author}
  {\bibfnamefont {M.}~\bibnamefont {Steinert}}, \bibinfo {author}
  {\bibfnamefont {C.}~\bibnamefont {Spielmann}}, \bibinfo {author}
  {\bibfnamefont {M.}~\bibnamefont {Z{\"{u}}rch}}, \bibinfo {author}
  {\bibfnamefont {T.}~\bibnamefont {Pertsch}}, \bibinfo {author} {\bibfnamefont
  {A.}~\bibnamefont {T{\"{u}}nnermann}}, \bibinfo {author} {\bibfnamefont
  {J.}~\bibnamefont {Limpert}}, \ and\ \bibinfo {author} {\bibfnamefont
  {J.}~\bibnamefont {Rothhardt}},\ }\href
  {https://www.osapublishing.org/abstract.cfm?URI=ol-41-22-5170} {\bibfield
  {journal} {\bibinfo  {journal} {Opt. Lett.}\ }\textbf {\bibinfo {volume}
  {41}},\ \bibinfo {pages} {5170} (\bibinfo {year} {2016})}\BibitemShut
  {NoStop}%
\bibitem [{\citenamefont {Fischer}\ \emph {et~al.}(2006)\citenamefont
  {Fischer}, \citenamefont {Kim}, \citenamefont {Chao}, \citenamefont {Liddle},
  \citenamefont {Anderson},\ and\ \citenamefont {Attwood}}]{Fischer2006}%
  \BibitemOpen
  \bibfield  {author} {\bibinfo {author} {\bibfnamefont {P.}~\bibnamefont
  {Fischer}}, \bibinfo {author} {\bibfnamefont {D.-H.}\ \bibnamefont {Kim}},
  \bibinfo {author} {\bibfnamefont {W.}~\bibnamefont {Chao}}, \bibinfo {author}
  {\bibfnamefont {J.~A.}\ \bibnamefont {Liddle}}, \bibinfo {author}
  {\bibfnamefont {E.~H.}\ \bibnamefont {Anderson}}, \ and\ \bibinfo {author}
  {\bibfnamefont {D.~T.}\ \bibnamefont {Attwood}},\ }\href
  {http://linkinghub.elsevier.com/retrieve/pii/S1369702105713353} {\bibfield
  {journal} {\bibinfo  {journal} {Mater. Today}\ }\textbf {\bibinfo {volume}
  {9}},\ \bibinfo {pages} {26} (\bibinfo {year} {2006})}\BibitemShut {NoStop}%
\bibitem [{\citenamefont {Kubo}\ \emph {et~al.}(2005)\citenamefont {Kubo},
  \citenamefont {Onda}, \citenamefont {Petek}, \citenamefont {Sun},
  \citenamefont {Jung},\ and\ \citenamefont {Kim}}]{Kubo2005}%
  \BibitemOpen
  \bibfield  {author} {\bibinfo {author} {\bibfnamefont {A.}~\bibnamefont
  {Kubo}}, \bibinfo {author} {\bibfnamefont {K.}~\bibnamefont {Onda}}, \bibinfo
  {author} {\bibfnamefont {H.}~\bibnamefont {Petek}}, \bibinfo {author}
  {\bibfnamefont {Z.}~\bibnamefont {Sun}}, \bibinfo {author} {\bibfnamefont
  {Y.~S.}\ \bibnamefont {Jung}}, \ and\ \bibinfo {author} {\bibfnamefont
  {H.~K.}\ \bibnamefont {Kim}},\ }\href
  {http://pubs.acs.org/doi/abs/10.1021/nl0506655} {\bibfield  {journal}
  {\bibinfo  {journal} {Nano Lett.}\ }\textbf {\bibinfo {volume} {5}},\
  \bibinfo {pages} {1123} (\bibinfo {year} {2005})}\BibitemShut {NoStop}%
\bibitem [{\citenamefont {{Meyer zu Heringdorf}}\ \emph
  {et~al.}(2007)\citenamefont {{Meyer zu Heringdorf}}, \citenamefont {Chelaru},
  \citenamefont {M{\"{o}}llenbeck}, \citenamefont {Thien},\ and\ \citenamefont
  {{Horn-von Hoegen}}}]{MeyerzuHeringdorf2007}%
  \BibitemOpen
  \bibfield  {author} {\bibinfo {author} {\bibfnamefont {F.-J.}\ \bibnamefont
  {{Meyer zu Heringdorf}}}, \bibinfo {author} {\bibfnamefont {L.}~\bibnamefont
  {Chelaru}}, \bibinfo {author} {\bibfnamefont {S.}~\bibnamefont
  {M{\"{o}}llenbeck}}, \bibinfo {author} {\bibfnamefont {D.}~\bibnamefont
  {Thien}}, \ and\ \bibinfo {author} {\bibfnamefont {M.}~\bibnamefont
  {{Horn-von Hoegen}}},\ }\href
  {http://linkinghub.elsevier.com/retrieve/pii/S0039602807006048} {\bibfield
  {journal} {\bibinfo  {journal} {Surf. Sci.}\ }\textbf {\bibinfo {volume}
  {601}},\ \bibinfo {pages} {4700} (\bibinfo {year} {2007})}\BibitemShut
  {NoStop}%
\bibitem [{\citenamefont {Man}\ \emph {et~al.}(2016)\citenamefont {Man},
  \citenamefont {Margiolakis}, \citenamefont {Deckoff-Jones}, \citenamefont
  {Harada}, \citenamefont {Wong}, \citenamefont {Krishna}, \citenamefont
  {Mad{\'{e}}o}, \citenamefont {Winchester}, \citenamefont {Lei}, \citenamefont
  {Vajtai}, \citenamefont {Ajayan},\ and\ \citenamefont {Dani}}]{Man2016}%
  \BibitemOpen
  \bibfield  {author} {\bibinfo {author} {\bibfnamefont {M.~K.~L.}\
  \bibnamefont {Man}}, \bibinfo {author} {\bibfnamefont {A.}~\bibnamefont
  {Margiolakis}}, \bibinfo {author} {\bibfnamefont {S.}~\bibnamefont
  {Deckoff-Jones}}, \bibinfo {author} {\bibfnamefont {T.}~\bibnamefont
  {Harada}}, \bibinfo {author} {\bibfnamefont {E.~L.}\ \bibnamefont {Wong}},
  \bibinfo {author} {\bibfnamefont {M.~B.~M.}\ \bibnamefont {Krishna}},
  \bibinfo {author} {\bibfnamefont {J.}~\bibnamefont {Mad{\'{e}}o}}, \bibinfo
  {author} {\bibfnamefont {A.}~\bibnamefont {Winchester}}, \bibinfo {author}
  {\bibfnamefont {S.}~\bibnamefont {Lei}}, \bibinfo {author} {\bibfnamefont
  {R.}~\bibnamefont {Vajtai}}, \bibinfo {author} {\bibfnamefont {P.~M.}\
  \bibnamefont {Ajayan}}, \ and\ \bibinfo {author} {\bibfnamefont {K.~M.}\
  \bibnamefont {Dani}},\ }\href
  {http://www.nature.com/doifinder/10.1038/nnano.2016.183} {\bibfield
  {journal} {\bibinfo  {journal} {Nat. Nanotechnol.}\ }\textbf {\bibinfo
  {volume} {12}},\ \bibinfo {pages} {36} (\bibinfo {year} {2016})}\BibitemShut
  {NoStop}%
\bibitem [{\citenamefont {D\c{a}browski}\ \emph {et~al.}(2016)\citenamefont
  {D\c{a}browski}, \citenamefont {Dai}, \citenamefont {Argondizzo},
  \citenamefont {Zou}, \citenamefont {Cui},\ and\ \citenamefont
  {Petek}}]{Dabrowski2016}%
  \BibitemOpen
  \bibfield  {author} {\bibinfo {author} {\bibfnamefont {M.}~\bibnamefont
  {D\c{a}browski}}, \bibinfo {author} {\bibfnamefont {Y.}~\bibnamefont {Dai}},
  \bibinfo {author} {\bibfnamefont {A.}~\bibnamefont {Argondizzo}}, \bibinfo
  {author} {\bibfnamefont {Q.}~\bibnamefont {Zou}}, \bibinfo {author}
  {\bibfnamefont {X.}~\bibnamefont {Cui}}, \ and\ \bibinfo {author}
  {\bibfnamefont {H.}~\bibnamefont {Petek}},\ }\href
  {http://pubs.acs.org/doi/10.1021/acsphotonics.6b00353} {\bibfield  {journal}
  {\bibinfo  {journal} {ACS Photonics}\ }\textbf {\bibinfo {volume} {3}},\
  \bibinfo {pages} {1704} (\bibinfo {year} {2016})}\BibitemShut {NoStop}%
\bibitem [{\citenamefont {Zewail}(2010)}]{uelec:Zewail}%
  \BibitemOpen
  \bibfield  {author} {\bibinfo {author} {\bibfnamefont {A.~H.}\ \bibnamefont
  {Zewail}},\ }\href
  {http://www.sciencemag.org/cgi/doi/10.1126/science.1166135} {\bibfield
  {journal} {\bibinfo  {journal} {Science}\ }\textbf {\bibinfo {volume}
  {328}},\ \bibinfo {pages} {187} (\bibinfo {year} {2010})}\BibitemShut
  {NoStop}%
\bibitem [{\citenamefont {Barwick}\ \emph {et~al.}(2009)\citenamefont
  {Barwick}, \citenamefont {Flannigan},\ and\ \citenamefont
  {Zewail}}]{Barwick2009}%
  \BibitemOpen
  \bibfield  {author} {\bibinfo {author} {\bibfnamefont {B.}~\bibnamefont
  {Barwick}}, \bibinfo {author} {\bibfnamefont {D.~J.}\ \bibnamefont
  {Flannigan}}, \ and\ \bibinfo {author} {\bibfnamefont {A.~H.}\ \bibnamefont
  {Zewail}},\ }\href {http://www.nature.com/doifinder/10.1038/nature08662}
  {\bibfield  {journal} {\bibinfo  {journal} {Nature}\ }\textbf {\bibinfo
  {volume} {462}},\ \bibinfo {pages} {902} (\bibinfo {year}
  {2009})}\BibitemShut {NoStop}%
\bibitem [{\citenamefont {Najafi}\ \emph {et~al.}(2015)\citenamefont {Najafi},
  \citenamefont {Scarborough}, \citenamefont {Tang},\ and\ \citenamefont
  {Zewail}}]{Najafi2015}%
  \BibitemOpen
  \bibfield  {author} {\bibinfo {author} {\bibfnamefont {E.}~\bibnamefont
  {Najafi}}, \bibinfo {author} {\bibfnamefont {T.~D.}\ \bibnamefont
  {Scarborough}}, \bibinfo {author} {\bibfnamefont {J.}~\bibnamefont {Tang}}, \
  and\ \bibinfo {author} {\bibfnamefont {A.}~\bibnamefont {Zewail}},\ }\href
  {http://www.sciencemag.org/cgi/doi/10.1126/science.aaa0217} {\bibfield
  {journal} {\bibinfo  {journal} {Science}\ }\textbf {\bibinfo {volume}
  {347}},\ \bibinfo {pages} {164} (\bibinfo {year} {2015})}\BibitemShut
  {NoStop}%
\bibitem [{\citenamefont {Feist}\ \emph {et~al.}(2017)\citenamefont {Feist},
  \citenamefont {Bach}, \citenamefont {{Rubiano da Silva}}, \citenamefont
  {Danz}, \citenamefont {M{\"{o}}ller}, \citenamefont {Priebe}, \citenamefont
  {Domr{\"{o}}se}, \citenamefont {Gatzmann}, \citenamefont {Rost},
  \citenamefont {Schauss}, \citenamefont {Strauch}, \citenamefont {Bormann},
  \citenamefont {Sivis}, \citenamefont {Sch{\"{a}}fer},\ and\ \citenamefont
  {Ropers}}]{Feist2016}%
  \BibitemOpen
  \bibfield  {author} {\bibinfo {author} {\bibfnamefont {A.}~\bibnamefont
  {Feist}}, \bibinfo {author} {\bibfnamefont {N.}~\bibnamefont {Bach}},
  \bibinfo {author} {\bibfnamefont {N.}~\bibnamefont {{Rubiano da Silva}}},
  \bibinfo {author} {\bibfnamefont {T.}~\bibnamefont {Danz}}, \bibinfo {author}
  {\bibfnamefont {M.}~\bibnamefont {M{\"{o}}ller}}, \bibinfo {author}
  {\bibfnamefont {K.~E.}\ \bibnamefont {Priebe}}, \bibinfo {author}
  {\bibfnamefont {T.}~\bibnamefont {Domr{\"{o}}se}}, \bibinfo {author}
  {\bibfnamefont {J.~G.}\ \bibnamefont {Gatzmann}}, \bibinfo {author}
  {\bibfnamefont {S.}~\bibnamefont {Rost}}, \bibinfo {author} {\bibfnamefont
  {J.}~\bibnamefont {Schauss}}, \bibinfo {author} {\bibfnamefont
  {S.}~\bibnamefont {Strauch}}, \bibinfo {author} {\bibfnamefont
  {R.}~\bibnamefont {Bormann}}, \bibinfo {author} {\bibfnamefont
  {M.}~\bibnamefont {Sivis}}, \bibinfo {author} {\bibfnamefont
  {S.}~\bibnamefont {Sch{\"{a}}fer}}, \ and\ \bibinfo {author} {\bibfnamefont
  {C.}~\bibnamefont {Ropers}},\ }\href {\doibase
  10.1016/j.ultramic.2016.12.005} {\bibfield  {journal} {\bibinfo  {journal}
  {Ultramicroscopy}\ }\textbf {\bibinfo {volume} {176}},\ \bibinfo {pages} {63}
  (\bibinfo {year} {2017})},\ \Eprint {http://arxiv.org/abs/1611.05022}
  {1611.05022} \BibitemShut {NoStop}%
\bibitem [{\citenamefont {Liao}\ \emph {et~al.}(2017)\citenamefont {Liao},
  \citenamefont {Zhao}, \citenamefont {Najafi}, \citenamefont {Yan},
  \citenamefont {Tian}, \citenamefont {Tice}, \citenamefont {Minnich},
  \citenamefont {Wang},\ and\ \citenamefont {Zewail}}]{Liao2017}%
  \BibitemOpen
  \bibfield  {author} {\bibinfo {author} {\bibfnamefont {B.}~\bibnamefont
  {Liao}}, \bibinfo {author} {\bibfnamefont {H.}~\bibnamefont {Zhao}}, \bibinfo
  {author} {\bibfnamefont {E.}~\bibnamefont {Najafi}}, \bibinfo {author}
  {\bibfnamefont {X.}~\bibnamefont {Yan}}, \bibinfo {author} {\bibfnamefont
  {H.}~\bibnamefont {Tian}}, \bibinfo {author} {\bibfnamefont {J.}~\bibnamefont
  {Tice}}, \bibinfo {author} {\bibfnamefont {A.~J.}\ \bibnamefont {Minnich}},
  \bibinfo {author} {\bibfnamefont {H.}~\bibnamefont {Wang}}, \ and\ \bibinfo
  {author} {\bibfnamefont {A.~H.}\ \bibnamefont {Zewail}},\ }\href
  {http://pubs.acs.org/doi/10.1021/acs.nanolett.7b00897} {\bibfield  {journal}
  {\bibinfo  {journal} {Nano Lett.}\ }\textbf {\bibinfo {volume} {17}},\
  \bibinfo {pages} {3675} (\bibinfo {year} {2017})}\BibitemShut {NoStop}%
\bibitem [{\citenamefont {da~Silva}\ \emph {et~al.}(2017)\citenamefont
  {da~Silva}, \citenamefont {M{\"{o}}ller}, \citenamefont {Feist},
  \citenamefont {Ulrichs}, \citenamefont {Ropers},\ and\ \citenamefont
  {Sch{\"{a}}fer}}]{DaSilva2017}%
  \BibitemOpen
  \bibfield  {author} {\bibinfo {author} {\bibfnamefont {N.~R.}\ \bibnamefont
  {da~Silva}}, \bibinfo {author} {\bibfnamefont {M.}~\bibnamefont
  {M{\"{o}}ller}}, \bibinfo {author} {\bibfnamefont {A.}~\bibnamefont {Feist}},
  \bibinfo {author} {\bibfnamefont {H.}~\bibnamefont {Ulrichs}}, \bibinfo
  {author} {\bibfnamefont {C.}~\bibnamefont {Ropers}}, \ and\ \bibinfo {author}
  {\bibfnamefont {S.}~\bibnamefont {Sch{\"{a}}fer}},\ }\href
  {http://arxiv.org/abs/1710.03307} {\  (\bibinfo {year} {2017})},\ \Eprint
  {http://arxiv.org/abs/1710.03307} {arXiv:1710.03307} \BibitemShut {NoStop}%
\bibitem [{\citenamefont {Priebe}\ \emph {et~al.}(2017)\citenamefont {Priebe},
  \citenamefont {Rathje}, \citenamefont {Yalunin}, \citenamefont {Hohage},
  \citenamefont {Feist}, \citenamefont {Sch{\"{a}}fer},\ and\ \citenamefont
  {Ropers}}]{Priebe2017}%
  \BibitemOpen
  \bibfield  {author} {\bibinfo {author} {\bibfnamefont {K.~E.}\ \bibnamefont
  {Priebe}}, \bibinfo {author} {\bibfnamefont {C.}~\bibnamefont {Rathje}},
  \bibinfo {author} {\bibfnamefont {S.~V.}\ \bibnamefont {Yalunin}}, \bibinfo
  {author} {\bibfnamefont {T.}~\bibnamefont {Hohage}}, \bibinfo {author}
  {\bibfnamefont {A.}~\bibnamefont {Feist}}, \bibinfo {author} {\bibfnamefont
  {S.}~\bibnamefont {Sch{\"{a}}fer}}, \ and\ \bibinfo {author} {\bibfnamefont
  {C.}~\bibnamefont {Ropers}},\ }\href
  {http://www.nature.com/articles/s41566-017-0045-8} {\bibfield  {journal}
  {\bibinfo  {journal} {Nat. Photonics}\ }\textbf {\bibinfo {volume} {11}},\
  \bibinfo {pages} {793} (\bibinfo {year} {2017})}\BibitemShut {NoStop}%
\bibitem [{\citenamefont {Morimoto}\ and\ \citenamefont
  {Baum}(2017)}]{Morimoto2017}%
  \BibitemOpen
  \bibfield  {author} {\bibinfo {author} {\bibfnamefont {Y.}~\bibnamefont
  {Morimoto}}\ and\ \bibinfo {author} {\bibfnamefont {P.}~\bibnamefont
  {Baum}},\ }\href {http://www.nature.com/articles/s41567-017-0007-6}
  {\bibfield  {journal} {\bibinfo  {journal} {Nat. Phys.}\ } (\bibinfo {year}
  {2017})}\BibitemShut {NoStop}%
\bibitem [{\citenamefont {Quinonez}\ \emph {et~al.}(2013)\citenamefont
  {Quinonez}, \citenamefont {Handali},\ and\ \citenamefont
  {Barwick}}]{Quinonez2013}%
  \BibitemOpen
  \bibfield  {author} {\bibinfo {author} {\bibfnamefont {E.}~\bibnamefont
  {Quinonez}}, \bibinfo {author} {\bibfnamefont {J.}~\bibnamefont {Handali}}, \
  and\ \bibinfo {author} {\bibfnamefont {B.}~\bibnamefont {Barwick}},\ }\href
  {http://aip.scitation.org/doi/10.1063/1.4827035} {\bibfield  {journal}
  {\bibinfo  {journal} {Rev. Sci. Instrum.}\ }\textbf {\bibinfo {volume}
  {84}},\ \bibinfo {pages} {103710} (\bibinfo {year} {2013})}\BibitemShut
  {NoStop}%
\bibitem [{\citenamefont {M{\"{u}}ller}\ \emph {et~al.}(2014)\citenamefont
  {M{\"{u}}ller}, \citenamefont {Paarmann},\ and\ \citenamefont
  {Ernstorfer}}]{Muller2014}%
  \BibitemOpen
  \bibfield  {author} {\bibinfo {author} {\bibfnamefont {M.}~\bibnamefont
  {M{\"{u}}ller}}, \bibinfo {author} {\bibfnamefont {A.}~\bibnamefont
  {Paarmann}}, \ and\ \bibinfo {author} {\bibfnamefont {R.}~\bibnamefont
  {Ernstorfer}},\ }\href@noop {} {\bibfield  {journal} {\bibinfo  {journal}
  {Nat. Commun.}\ }\textbf {\bibinfo {volume} {5}},\ \bibinfo {pages} {5292}
  (\bibinfo {year} {2014})}\BibitemShut {NoStop}%
\bibitem [{\citenamefont {Bainbridge}\ \emph {et~al.}(2016)\citenamefont
  {Bainbridge}, \citenamefont {{Barlow Myers}},\ and\ \citenamefont
  {Bryan}}]{Bainbridge2016}%
  \BibitemOpen
  \bibfield  {author} {\bibinfo {author} {\bibfnamefont {A.~R.}\ \bibnamefont
  {Bainbridge}}, \bibinfo {author} {\bibfnamefont {C.~W.}\ \bibnamefont
  {{Barlow Myers}}}, \ and\ \bibinfo {author} {\bibfnamefont {W.~A.}\
  \bibnamefont {Bryan}},\ }\href
  {http://aca.scitation.org/doi/10.1063/1.4947098} {\bibfield  {journal}
  {\bibinfo  {journal} {Struct. Dyn.}\ }\textbf {\bibinfo {volume} {3}},\
  \bibinfo {pages} {023612} (\bibinfo {year} {2016})}\BibitemShut {NoStop}%
\bibitem [{\citenamefont {Longchamp}\ \emph {et~al.}(2017)\citenamefont
  {Longchamp}, \citenamefont {Rauschenbach}, \citenamefont {Abb}, \citenamefont
  {Escher}, \citenamefont {Latychevskaia}, \citenamefont {Kern},\ and\
  \citenamefont {Fink}}]{Longchamp2015a}%
  \BibitemOpen
  \bibfield  {author} {\bibinfo {author} {\bibfnamefont {J.-N.}\ \bibnamefont
  {Longchamp}}, \bibinfo {author} {\bibfnamefont {S.}~\bibnamefont
  {Rauschenbach}}, \bibinfo {author} {\bibfnamefont {S.}~\bibnamefont {Abb}},
  \bibinfo {author} {\bibfnamefont {C.}~\bibnamefont {Escher}}, \bibinfo
  {author} {\bibfnamefont {T.}~\bibnamefont {Latychevskaia}}, \bibinfo {author}
  {\bibfnamefont {K.}~\bibnamefont {Kern}}, \ and\ \bibinfo {author}
  {\bibfnamefont {H.-W.}\ \bibnamefont {Fink}},\ }\href@noop {} {\bibfield
  {journal} {\bibinfo  {journal} {Proc. Natl. Acad. Sci.}\ }\textbf {\bibinfo
  {volume} {114}},\ \bibinfo {pages} {1474} (\bibinfo {year} {2017})},\ \Eprint
  {http://arxiv.org/abs/1512.08958} {1512.08958} \BibitemShut {NoStop}%
\bibitem [{\citenamefont {Paarmann}\ \emph {et~al.}(2012)\citenamefont
  {Paarmann}, \citenamefont {Gulde}, \citenamefont {M{\"{u}}ller},
  \citenamefont {Sch{\"{a}}fer}, \citenamefont {Schweda}, \citenamefont
  {Maiti}, \citenamefont {Xu}, \citenamefont {Hohage}, \citenamefont {Schenk},
  \citenamefont {Ropers},\ and\ \citenamefont {Ernstorfer}}]{Paarmann2012}%
  \BibitemOpen
  \bibfield  {author} {\bibinfo {author} {\bibfnamefont {A.}~\bibnamefont
  {Paarmann}}, \bibinfo {author} {\bibfnamefont {M.}~\bibnamefont {Gulde}},
  \bibinfo {author} {\bibfnamefont {M.}~\bibnamefont {M{\"{u}}ller}}, \bibinfo
  {author} {\bibfnamefont {S.}~\bibnamefont {Sch{\"{a}}fer}}, \bibinfo {author}
  {\bibfnamefont {S.}~\bibnamefont {Schweda}}, \bibinfo {author} {\bibfnamefont
  {M.}~\bibnamefont {Maiti}}, \bibinfo {author} {\bibfnamefont
  {C.}~\bibnamefont {Xu}}, \bibinfo {author} {\bibfnamefont {T.}~\bibnamefont
  {Hohage}}, \bibinfo {author} {\bibfnamefont {F.}~\bibnamefont {Schenk}},
  \bibinfo {author} {\bibfnamefont {C.}~\bibnamefont {Ropers}}, \ and\ \bibinfo
  {author} {\bibfnamefont {R.}~\bibnamefont {Ernstorfer}},\ }\href@noop {}
  {\bibfield  {journal} {\bibinfo  {journal} {J. Appl. Phys.}\ }\textbf
  {\bibinfo {volume} {112}},\ \bibinfo {pages} {113109} (\bibinfo {year}
  {2012})}\BibitemShut {NoStop}%
\bibitem [{\citenamefont {Vogelsang}\ \emph {et~al.}(2015)\citenamefont
  {Vogelsang}, \citenamefont {Robin}, \citenamefont {Nagy}, \citenamefont
  {Dombi}, \citenamefont {Rosenkranz}, \citenamefont {Schiek}, \citenamefont
  {Gro{\ss}},\ and\ \citenamefont {Lienau}}]{Vogelsang2015}%
  \BibitemOpen
  \bibfield  {author} {\bibinfo {author} {\bibfnamefont {J.}~\bibnamefont
  {Vogelsang}}, \bibinfo {author} {\bibfnamefont {J.}~\bibnamefont {Robin}},
  \bibinfo {author} {\bibfnamefont {B.~J.}\ \bibnamefont {Nagy}}, \bibinfo
  {author} {\bibfnamefont {P.}~\bibnamefont {Dombi}}, \bibinfo {author}
  {\bibfnamefont {D.}~\bibnamefont {Rosenkranz}}, \bibinfo {author}
  {\bibfnamefont {M.}~\bibnamefont {Schiek}}, \bibinfo {author} {\bibfnamefont
  {P.}~\bibnamefont {Gro{\ss}}}, \ and\ \bibinfo {author} {\bibfnamefont
  {C.}~\bibnamefont {Lienau}},\ }\href@noop {} {\bibfield  {journal} {\bibinfo
  {journal} {Nano Lett.}\ }\textbf {\bibinfo {volume} {15}},\ \bibinfo {pages}
  {4685} (\bibinfo {year} {2015})}\BibitemShut {NoStop}%
\bibitem [{\citenamefont {M{\"{u}}ller}\ \emph {et~al.}(2016)\citenamefont
  {M{\"{u}}ller}, \citenamefont {Kravtsov}, \citenamefont {Paarmann},
  \citenamefont {Raschke},\ and\ \citenamefont {Ernstorfer}}]{Muller2016a}%
  \BibitemOpen
  \bibfield  {author} {\bibinfo {author} {\bibfnamefont {M.}~\bibnamefont
  {M{\"{u}}ller}}, \bibinfo {author} {\bibfnamefont {V.}~\bibnamefont
  {Kravtsov}}, \bibinfo {author} {\bibfnamefont {A.}~\bibnamefont {Paarmann}},
  \bibinfo {author} {\bibfnamefont {M.~B.}\ \bibnamefont {Raschke}}, \ and\
  \bibinfo {author} {\bibfnamefont {R.}~\bibnamefont {Ernstorfer}},\ }\href
  {http://pubs.acs.org/doi/abs/10.1021/acsphotonics.5b00710} {\bibfield
  {journal} {\bibinfo  {journal} {ACS Photonics}\ }\textbf {\bibinfo {volume}
  {3}},\ \bibinfo {pages} {611} (\bibinfo {year} {2016})}\BibitemShut {NoStop}%
\bibitem [{\citenamefont {Beyer}\ and\ \citenamefont
  {G{\"{o}}lzh{\"{a}}user}(2010)}]{Beyer2010}%
  \BibitemOpen
  \bibfield  {author} {\bibinfo {author} {\bibfnamefont {A.}~\bibnamefont
  {Beyer}}\ and\ \bibinfo {author} {\bibfnamefont {A.}~\bibnamefont
  {G{\"{o}}lzh{\"{a}}user}},\ }\href@noop {} {\bibfield  {journal} {\bibinfo
  {journal} {J. Phys. Condens. Matter}\ }\textbf {\bibinfo {volume} {22}},\
  \bibinfo {pages} {343001} (\bibinfo {year} {2010})}\BibitemShut {NoStop}%
\bibitem [{\citenamefont {Morton}\ and\ \citenamefont
  {Ramberg}(1939)}]{Morton1939}%
  \BibitemOpen
  \bibfield  {author} {\bibinfo {author} {\bibfnamefont {G.~A.}\ \bibnamefont
  {Morton}}\ and\ \bibinfo {author} {\bibfnamefont {E.~G.}\ \bibnamefont
  {Ramberg}},\ }\href {https://link.aps.org/doi/10.1103/PhysRev.56.705}
  {\bibfield  {journal} {\bibinfo  {journal} {Phys. Rev.}\ }\textbf {\bibinfo
  {volume} {56}},\ \bibinfo {pages} {705} (\bibinfo {year} {1939})}\BibitemShut
  {NoStop}%
\bibitem [{\citenamefont {Spence}\ \emph {et~al.}(1993)\citenamefont {Spence},
  \citenamefont {Qian},\ and\ \citenamefont {Melmed}}]{Spence1993}%
  \BibitemOpen
  \bibfield  {author} {\bibinfo {author} {\bibfnamefont {J.}~\bibnamefont
  {Spence}}, \bibinfo {author} {\bibfnamefont {W.}~\bibnamefont {Qian}}, \ and\
  \bibinfo {author} {\bibfnamefont {A.}~\bibnamefont {Melmed}},\ }\href
  {http://linkinghub.elsevier.com/retrieve/pii/0304399193900634} {\bibfield
  {journal} {\bibinfo  {journal} {Ultramicroscopy}\ }\textbf {\bibinfo {volume}
  {52}},\ \bibinfo {pages} {473} (\bibinfo {year} {1993})}\BibitemShut
  {NoStop}%
\bibitem [{\citenamefont {Lai}\ \emph {et~al.}(1999)\citenamefont {Lai},
  \citenamefont {Degiovanni},\ and\ \citenamefont {Morin}}]{Lai1999}%
  \BibitemOpen
  \bibfield  {author} {\bibinfo {author} {\bibfnamefont {W.}~\bibnamefont
  {Lai}}, \bibinfo {author} {\bibfnamefont {A.}~\bibnamefont {Degiovanni}}, \
  and\ \bibinfo {author} {\bibfnamefont {R.}~\bibnamefont {Morin}},\
  }\href@noop {} {\bibfield  {journal} {\bibinfo  {journal} {Appl. Phys.
  Lett.}\ }\textbf {\bibinfo {volume} {74}},\ \bibinfo {pages} {618} (\bibinfo
  {year} {1999})}\BibitemShut {NoStop}%
\bibitem [{\citenamefont {Dolocan}\ \emph {et~al.}(2006)\citenamefont
  {Dolocan}, \citenamefont {Hengsberger}, \citenamefont {Neff}, \citenamefont
  {Barry}, \citenamefont {Cirelli}, \citenamefont {Greber},\ and\ \citenamefont
  {Osterwalder}}]{Dolocan2006}%
  \BibitemOpen
  \bibfield  {author} {\bibinfo {author} {\bibfnamefont {A.}~\bibnamefont
  {Dolocan}}, \bibinfo {author} {\bibfnamefont {M.}~\bibnamefont
  {Hengsberger}}, \bibinfo {author} {\bibfnamefont {H.~J.}\ \bibnamefont
  {Neff}}, \bibinfo {author} {\bibfnamefont {M.}~\bibnamefont {Barry}},
  \bibinfo {author} {\bibfnamefont {C.}~\bibnamefont {Cirelli}}, \bibinfo
  {author} {\bibfnamefont {T.}~\bibnamefont {Greber}}, \ and\ \bibinfo {author}
  {\bibfnamefont {J.}~\bibnamefont {Osterwalder}},\ }\href
  {http://stacks.iop.org/1347-4065/45/285} {\bibfield  {journal} {\bibinfo
  {journal} {Japanese Journal of Applied Physics}\ }\textbf {\bibinfo {volume}
  {45}},\ \bibinfo {pages} {285} (\bibinfo {year} {2006})}\BibitemShut
  {NoStop}%
\bibitem [{\citenamefont {Dwyer}\ \emph {et~al.}(2006)\citenamefont {Dwyer},
  \citenamefont {Hebeisen}, \citenamefont {Ernstorfer}, \citenamefont {Harb},
  \citenamefont {B.Deyirmenjian}, \citenamefont {Jordan},\ and\ \citenamefont
  {Miller}}]{uelec:Dwyer}%
  \BibitemOpen
  \bibfield  {author} {\bibinfo {author} {\bibfnamefont {J.~R.}\ \bibnamefont
  {Dwyer}}, \bibinfo {author} {\bibfnamefont {C.~T.}\ \bibnamefont {Hebeisen}},
  \bibinfo {author} {\bibfnamefont {R.}~\bibnamefont {Ernstorfer}}, \bibinfo
  {author} {\bibfnamefont {M.}~\bibnamefont {Harb}}, \bibinfo {author}
  {\bibfnamefont {V.}~\bibnamefont {B.Deyirmenjian}}, \bibinfo {author}
  {\bibfnamefont {R.~E.}\ \bibnamefont {Jordan}}, \ and\ \bibinfo {author}
  {\bibfnamefont {R.~J.~D.}\ \bibnamefont {Miller}},\ }\href@noop {} {\bibfield
   {journal} {\bibinfo  {journal} {Phil. trans. R. Soc. A}\ }\textbf {\bibinfo
  {volume} {364}},\ \bibinfo {pages} {741} (\bibinfo {year}
  {2006})}\BibitemShut {NoStop}%
\bibitem [{\citenamefont {Hebeisen}\ \emph {et~al.}(2008)\citenamefont
  {Hebeisen}, \citenamefont {Sciaini}, \citenamefont {Harb}, \citenamefont
  {Ernstorfer}, \citenamefont {Kruglik},\ and\ \citenamefont
  {Miller}}]{Hebeisen2008}%
  \BibitemOpen
  \bibfield  {author} {\bibinfo {author} {\bibfnamefont {C.~T.}\ \bibnamefont
  {Hebeisen}}, \bibinfo {author} {\bibfnamefont {G.}~\bibnamefont {Sciaini}},
  \bibinfo {author} {\bibfnamefont {M.}~\bibnamefont {Harb}}, \bibinfo {author}
  {\bibfnamefont {R.}~\bibnamefont {Ernstorfer}}, \bibinfo {author}
  {\bibfnamefont {S.~G.}\ \bibnamefont {Kruglik}}, \ and\ \bibinfo {author}
  {\bibfnamefont {R.~J.~D.}\ \bibnamefont {Miller}},\ }\href
  {https://link.aps.org/doi/10.1103/PhysRevB.78.081403} {\bibfield  {journal}
  {\bibinfo  {journal} {Phys. Rev. B}\ }\textbf {\bibinfo {volume} {78}},\
  \bibinfo {pages} {081403} (\bibinfo {year} {2008})}\BibitemShut {NoStop}%
\bibitem [{\citenamefont {Centurion}\ \emph {et~al.}(2008)\citenamefont
  {Centurion}, \citenamefont {Reckenthaeler}, \citenamefont {Trushin},
  \citenamefont {Krausz},\ and\ \citenamefont {Fill}}]{Centurion2008}%
  \BibitemOpen
  \bibfield  {author} {\bibinfo {author} {\bibfnamefont {M.}~\bibnamefont
  {Centurion}}, \bibinfo {author} {\bibfnamefont {P.}~\bibnamefont
  {Reckenthaeler}}, \bibinfo {author} {\bibfnamefont {S.~A.}\ \bibnamefont
  {Trushin}}, \bibinfo {author} {\bibfnamefont {F.}~\bibnamefont {Krausz}}, \
  and\ \bibinfo {author} {\bibfnamefont {E.~E.}\ \bibnamefont {Fill}},\ }\href
  {\doibase 10.1038/nphoton.2008.77} {\bibfield  {journal} {\bibinfo  {journal}
  {Nat. Photonics}\ }\textbf {\bibinfo {volume} {2}},\ \bibinfo {pages} {315}
  (\bibinfo {year} {2008})}\BibitemShut {NoStop}%
\bibitem [{\citenamefont {Yanagisawa}\ \emph {et~al.}(2011)\citenamefont
  {Yanagisawa}, \citenamefont {Hengsberger}, \citenamefont {Leuenberger},
  \citenamefont {Kl{\"{o}}ckner}, \citenamefont {Hafner}, \citenamefont
  {Greber},\ and\ \citenamefont {Osterwalder}}]{Yanagisawa2011}%
  \BibitemOpen
  \bibfield  {author} {\bibinfo {author} {\bibfnamefont {H.}~\bibnamefont
  {Yanagisawa}}, \bibinfo {author} {\bibfnamefont {M.}~\bibnamefont
  {Hengsberger}}, \bibinfo {author} {\bibfnamefont {D.}~\bibnamefont
  {Leuenberger}}, \bibinfo {author} {\bibfnamefont {M.}~\bibnamefont
  {Kl{\"{o}}ckner}}, \bibinfo {author} {\bibfnamefont {C.}~\bibnamefont
  {Hafner}}, \bibinfo {author} {\bibfnamefont {T.}~\bibnamefont {Greber}}, \
  and\ \bibinfo {author} {\bibfnamefont {J.}~\bibnamefont {Osterwalder}},\
  }\href {https://link.aps.org/doi/10.1103/PhysRevLett.107.087601} {\bibfield
  {journal} {\bibinfo  {journal} {Phys. Rev. Lett.}\ }\textbf {\bibinfo
  {volume} {107}},\ \bibinfo {pages} {087601} (\bibinfo {year}
  {2011})}\BibitemShut {NoStop}%
\bibitem [{\citenamefont {Latychevskaia}\ \emph {et~al.}(2016)\citenamefont
  {Latychevskaia}, \citenamefont {Wicki}, \citenamefont {Longchamp},
  \citenamefont {Escher},\ and\ \citenamefont {Fink}}]{Latychevskaia2016}%
  \BibitemOpen
  \bibfield  {author} {\bibinfo {author} {\bibfnamefont {T.}~\bibnamefont
  {Latychevskaia}}, \bibinfo {author} {\bibfnamefont {F.}~\bibnamefont
  {Wicki}}, \bibinfo {author} {\bibfnamefont {J.-N.}\ \bibnamefont
  {Longchamp}}, \bibinfo {author} {\bibfnamefont {C.}~\bibnamefont {Escher}}, \
  and\ \bibinfo {author} {\bibfnamefont {H.-W.}\ \bibnamefont {Fink}},\ }\href
  {http://pubs.acs.org/doi/abs/10.1021/acs.nanolett.6b01881} {\bibfield
  {journal} {\bibinfo  {journal} {Nano Lett.}\ }\textbf {\bibinfo {volume}
  {16}},\ \bibinfo {pages} {5469} (\bibinfo {year} {2016})}\BibitemShut
  {NoStop}%
\bibitem [{\citenamefont {Yanagisawa}\ \emph {et~al.}(2016)\citenamefont
  {Yanagisawa}, \citenamefont {Schnepp}, \citenamefont {Hafner}, \citenamefont
  {Hengsberger}, \citenamefont {Kim}, \citenamefont {Kling}, \citenamefont
  {Landsman}, \citenamefont {Gallmann},\ and\ \citenamefont
  {Osterwalder}}]{Yanagisawa2016}%
  \BibitemOpen
  \bibfield  {author} {\bibinfo {author} {\bibfnamefont {H.}~\bibnamefont
  {Yanagisawa}}, \bibinfo {author} {\bibfnamefont {S.}~\bibnamefont {Schnepp}},
  \bibinfo {author} {\bibfnamefont {C.}~\bibnamefont {Hafner}}, \bibinfo
  {author} {\bibfnamefont {M.}~\bibnamefont {Hengsberger}}, \bibinfo {author}
  {\bibfnamefont {D.~E.}\ \bibnamefont {Kim}}, \bibinfo {author} {\bibfnamefont
  {M.~F.}\ \bibnamefont {Kling}}, \bibinfo {author} {\bibfnamefont
  {A.}~\bibnamefont {Landsman}}, \bibinfo {author} {\bibfnamefont
  {L.}~\bibnamefont {Gallmann}}, \ and\ \bibinfo {author} {\bibfnamefont
  {J.}~\bibnamefont {Osterwalder}},\ }\href
  {http://www.nature.com/articles/srep35877} {\bibfield  {journal} {\bibinfo
  {journal} {Scientific Reports}\ }\textbf {\bibinfo {volume} {6}},\ \bibinfo
  {pages} {35877} (\bibinfo {year} {2016})}\BibitemShut {NoStop}%
\bibitem [{\citenamefont {Kinzel}\ \emph {et~al.}(2016)\citenamefont {Kinzel},
  \citenamefont {Sch{\"{u}}lein}, \citenamefont {Wei{\ss}}, \citenamefont
  {Janker}, \citenamefont {B{\"{u}}hler}, \citenamefont {Heigl}, \citenamefont
  {Rudolph}, \citenamefont {Mork{\"{o}}tter}, \citenamefont {D{\"{o}}blinger},
  \citenamefont {Bichler}, \citenamefont {Abstreiter}, \citenamefont {Finley},
  \citenamefont {Wixforth}, \citenamefont {Koblm{\"{u}}ller},\ and\
  \citenamefont {Krenner}}]{Kinzel2016}%
  \BibitemOpen
  \bibfield  {author} {\bibinfo {author} {\bibfnamefont {J.~B.}\ \bibnamefont
  {Kinzel}}, \bibinfo {author} {\bibfnamefont {F.~J.~R.}\ \bibnamefont
  {Sch{\"{u}}lein}}, \bibinfo {author} {\bibfnamefont {M.}~\bibnamefont
  {Wei{\ss}}}, \bibinfo {author} {\bibfnamefont {L.}~\bibnamefont {Janker}},
  \bibinfo {author} {\bibfnamefont {D.~D.}\ \bibnamefont {B{\"{u}}hler}},
  \bibinfo {author} {\bibfnamefont {M.}~\bibnamefont {Heigl}}, \bibinfo
  {author} {\bibfnamefont {D.}~\bibnamefont {Rudolph}}, \bibinfo {author}
  {\bibfnamefont {S.}~\bibnamefont {Mork{\"{o}}tter}}, \bibinfo {author}
  {\bibfnamefont {M.}~\bibnamefont {D{\"{o}}blinger}}, \bibinfo {author}
  {\bibfnamefont {M.}~\bibnamefont {Bichler}}, \bibinfo {author} {\bibfnamefont
  {G.}~\bibnamefont {Abstreiter}}, \bibinfo {author} {\bibfnamefont {J.~J.}\
  \bibnamefont {Finley}}, \bibinfo {author} {\bibfnamefont {A.}~\bibnamefont
  {Wixforth}}, \bibinfo {author} {\bibfnamefont {G.}~\bibnamefont
  {Koblm{\"{u}}ller}}, \ and\ \bibinfo {author} {\bibfnamefont {H.~J.}\
  \bibnamefont {Krenner}},\ }\href
  {http://pubs.acs.org/doi/abs/10.1021/acsnano.5b07639} {\bibfield  {journal}
  {\bibinfo  {journal} {ACS Nano}\ }\textbf {\bibinfo {volume} {10}},\ \bibinfo
  {pages} {4942} (\bibinfo {year} {2016})}\BibitemShut {NoStop}%
\bibitem [{\citenamefont {Prechtel}\ \emph {et~al.}(2011)\citenamefont
  {Prechtel}, \citenamefont {Song}, \citenamefont {Manus}, \citenamefont
  {Schuh}, \citenamefont {Wegscheider},\ and\ \citenamefont
  {Holleitner}}]{Prechtel2011}%
  \BibitemOpen
  \bibfield  {author} {\bibinfo {author} {\bibfnamefont {L.}~\bibnamefont
  {Prechtel}}, \bibinfo {author} {\bibfnamefont {L.}~\bibnamefont {Song}},
  \bibinfo {author} {\bibfnamefont {S.}~\bibnamefont {Manus}}, \bibinfo
  {author} {\bibfnamefont {D.}~\bibnamefont {Schuh}}, \bibinfo {author}
  {\bibfnamefont {W.}~\bibnamefont {Wegscheider}}, \ and\ \bibinfo {author}
  {\bibfnamefont {A.~W.}\ \bibnamefont {Holleitner}},\ }\href
  {http://www.ncbi.nlm.nih.gov/pubmed/21142051
  http://pubs.acs.org/doi/abs/10.1021/nl1036897} {\bibfield  {journal}
  {\bibinfo  {journal} {Nano Lett.}\ }\textbf {\bibinfo {volume} {11}},\
  \bibinfo {pages} {269} (\bibinfo {year} {2011})}\BibitemShut {NoStop}%
\bibitem [{\citenamefont {Hong}\ \emph {et~al.}(2014)\citenamefont {Hong},
  \citenamefont {Kim}, \citenamefont {Shi}, \citenamefont {Zhang},
  \citenamefont {Jin}, \citenamefont {Sun}, \citenamefont {Tongay},
  \citenamefont {Wu}, \citenamefont {Zhang},\ and\ \citenamefont
  {Wang}}]{Hong2014}%
  \BibitemOpen
  \bibfield  {author} {\bibinfo {author} {\bibfnamefont {X.}~\bibnamefont
  {Hong}}, \bibinfo {author} {\bibfnamefont {J.}~\bibnamefont {Kim}}, \bibinfo
  {author} {\bibfnamefont {S.-F.}\ \bibnamefont {Shi}}, \bibinfo {author}
  {\bibfnamefont {Y.}~\bibnamefont {Zhang}}, \bibinfo {author} {\bibfnamefont
  {C.}~\bibnamefont {Jin}}, \bibinfo {author} {\bibfnamefont {Y.}~\bibnamefont
  {Sun}}, \bibinfo {author} {\bibfnamefont {S.}~\bibnamefont {Tongay}},
  \bibinfo {author} {\bibfnamefont {J.}~\bibnamefont {Wu}}, \bibinfo {author}
  {\bibfnamefont {Y.}~\bibnamefont {Zhang}}, \ and\ \bibinfo {author}
  {\bibfnamefont {F.}~\bibnamefont {Wang}},\ }\href
  {http://www.nature.com/doifinder/10.1038/nnano.2014.167} {\bibfield
  {journal} {\bibinfo  {journal} {Nature Nanotechnology}\ }\textbf {\bibinfo
  {volume} {9}},\ \bibinfo {pages} {682} (\bibinfo {year} {2014})}\BibitemShut
  {NoStop}%
\end{thebibliography}%

\end{document}